\documentclass[12pt]{elsart}
\usepackage{amsmath,amssymb}
\usepackage{graphicx,psfrag,here,epsfig}
\usepackage{pstricks}
\usepackage{pst-node}
\usepackage{epsf}
\usepackage{citesort}
\allowdisplaybreaks[1]

\begin{document}
\newcommand{\co}{\; \; ,}
\def\words#1{\mbox{\small{\,#1}}}
\def\bea{\begin{eqnarray}}
\def\eea{\end{eqnarray}}
\def\eq{\begin{eqnarray}}
\def\en{\end{eqnarray}}
\def\be{\begin{equation}}
\def\ee{\end{equation}}
\newcommand{\ed}{\end{document}}
\newcommand{\rr}{\mbox{\boldmath $r$}}
\newcommand{\rrn}{\mbox{$r$}}
\newcommand{\rp}{\mbox{\boldmath $p$}}

\newcommand{\nnnl}{\nonumber\\}
\newcommand{\fs}{\, . \,}
\def\query#1{\marginpar{\begin{flushleft}\footnotesize#1\end{flushleft}}}%

\runauthor{Gay Ducati, Machado and Machado}
\renewcommand{\theequation}{\arabic{equation}}
\begin{frontmatter}

\title{\Large\bf Neutral current neutrino-nucleus interactions at high energies}

\author[UFRGS]{M.B.~Gay Ducati,}
\author[UFRGS]{M.M.~Machado,}
\author[UNIPAMPA]{M.V.T.~Machado,}

\address[UFRGS]{High Energy Physics Phenomenology Group, GFPAE,  IF-UFRGS \\
Caixa Postal 15051, CEP 91501-970, Porto Alegre, RS, Brazil}
\address[UNIPAMPA]{Centro de Ci\^encias Exatas e Tecnol\'ogicas, Universidade Federal do Pampa \\
Campus de Bag\'e, Rua Carlos Barbosa. CEP 96400-970. Bag\'e, RS, Brazil}

\begin{abstract}
  We present a QCD analysis of the neutral current neutrino-nucleus interaction at the small-$x$ region using the color dipole formalism. This phenomenological approach is quite successful in describing experimental results in deep inelastic $ep$ scattering and charged current neutrino-nucleus interactions at high energies. We present theoretical predictions for the relevant structure functions and the corresponding implications for the total NC neutrino cross section.

\end{abstract}


\end{frontmatter}

\section{Introduction}

The interaction of  high energy neutrinos in hadron targets are an important probe to test Quantum Chromodynamics (QCD) and useful to understand the parton properties of hadron structure. The several combinations of neutrino and anti-neutrino scattering data can be used to determine the structure functions, which constrain the valence, sea and gluon parton distributions in the nucleons or nuclei. In a similar way as for charged-lepton deep inelastic scattering (DIS), the deep inelastic neutrino scattering is also used to investigate the structure of nucleons and nuclei. In the leading order quark-parton model, the structure function $F_2$ is the singlet distribution, $F_2^{\nu N}\propto xq^S=x\sum(q+\bar{q})$, the sum of momentum densities of all interacting quarks constituents. These relations are further modified by higher-order QCD corrections.  Currently, the theoretical description of experimental data in neutrino DIS is reasonable, where main theoretical uncertainties are the role played by nuclear shadowing in contrast with lepton-charged DIS, and a correct understanding of the low $Q^2$ limit. However, nuclear effects are taken into account by using the nuclear ratios $R=F_2^A/AF_2^p$ extracted from lepton-nucleus DIS, which  could be different for the neutrino-nucleus case. The low-$Q^2$ region cannot be addressed within the pQCD quark-parton model since a hard momentum scale $Q_0^2\geq 1-2$ GeV$^2$ is required in order to perform perturbative expansion.

Concerning the important issues raised above, in Ref. \cite{gmm} we have performed an analysis of small-$x$ charged current (CC) neutrino-nucleus DIS using the color dipole formalism \cite{DIPOLEPIC}. The structure functions $F_2^{\nu N}$,  $xF_3^{\nu N}$ and the quantity $\Delta xF_3^{\nu N}$ were calculated and compared with the experimental data from CCFR \cite{CCFR,Fleming,CCFR3} and NuTeV \cite{Tzanov} by employing phenomenological parameterizations for the dipole cross section which successfully describe small-$x$ inclusive and diffractive $ep$ DIS data. In addition, we predicted the nuclear ratios $R_2$ and $R_3$ and single out the size of nuclear effects in each case. It was found that small-$x$ data show geometric scaling property for the boson-hadron cross section and the structure function $F_2$ is in agreement with the phenomenological implementation using the saturation models at the small-$x$ region \cite{gmm}.  The structure functions $xF_3^{\nu N}$ was discussed in detail, where the sea content described by the quantity $\Delta xF_3^{\nu N}$ is well described.  This investigation calls attention to the robustness of the  color dipole formalism to describe the total neutrino cross section of ultra-high energy neutrinos at very high energies.

Our goal in this work is to perform similar investigation for neutrino neutral current (NC) scattering, which is an ideal probe for new physics. As an additional motivation to study NC neutrino interactions we quote the purpose of a new high-energy,
  ultra-high statistics neutrino scattering experiment, NuSOnG \cite{nusong}
  (Neutrino Scattering on Glass).  This experiment uses a
  Tevatron-based neutrino beam to obtain  an order of magnitude
  higher statistics than presently available for the purely weak $\nu_{\mu}+e^-$ interactions and DIS  events which is about two orders of magnitude larger than
  past measurements will also be obtained.  The very high statistics will permit
an electroweak measurement using the DIS data sample from NuSOnG, through the ``Paschos Wolfenstein method'' (PW) \cite{PW}.  The best electroweak measurement using DIS events to date
comes from the NuTeV experiment, which has observed an anomaly.  The
  DIS measurement requires the knowledge of PDFs which describe the
  momentum distribution of quarks as a function of $Q^2$ and can
bring in theoretical uncertainties from sources such as nuclear
effects and nuclear isospin violation. NuSOnG probably will improve statistics in comparison to previous
highest statistics experiments NuTeV and CCFR. Theoretical uncertainties can be addressed by making a high statistics measurement of the PDFs on glass, {\it in situ}. NuSonG  will
generate an unprecedent sample of $>100$M DIS events which can be
used to measure six structure functions (three on neutrinos and three
for anti-neutrinos) as well as the strange and anti-strange parton
distributions. The NuSOnG experiment intends to  measure separately  $\Delta F_{2}\equiv\frac{5}{18}\, F_{2}^{CC}(x,Q^{2})-F_{2}^{NC}(x,Q^{2})$
in CC $W^{\pm}$ exchange and NC $\gamma/Z$ exchange processes, which can help to constrain charge symmetry violation.

In this work we present a determination of the small-$x$ structure
functions for NC neutrino-nucleus within the color dipole
formalism. This approach allows for a simple implementation of
shadowing corrections \cite{ARMESTO} in neutrino-nuclei
interactions. This paper is organized as follows. In Sec. 2, the
structure functions $F_2^{\nu N}$,  $F_L^{\nu N}$ and the nuclear
ratio  $R_A= F_2^{\nu A}/AF_2^{\nu N}$ are investigated in this
approach at small-$x$ region, employing recent
phenomenological parton saturation models.  It is shown that at
small-$x$, the NC boson-nucleon cross section should exhibit the geometric scaling property, that  has important consequences for ultra-high energy neutrino phenomenology. Finally, we also analyze the charm contribution to the total NC neutrino-nucleus cross section. The prediction is consistent with the existing experimental determinations. In the last section we summarize the results and present our conclusions.

\section{Neutral current neutrino-nucleon interaction at high energies}

In the color dipole approach \cite{DIPOLEPIC}, the small-$x$ deep
inelastic scattering (DIS) is treated in terms of the interaction of
the $q\bar{q}$ color dipole of size $r$ with the target nucleon (or
nucleus) which is characterized by the beam/flavor independent color
dipole cross section. The color dipole formalism allows an all twist
computation, in contrast with the usual leading twist approximation of
the structure functions. In the neutral current neutrino interaction,
the process $Z^0+N\rightarrow X$ is viewed as the result of the
interaction of a color singlet  dipole, in which the gauge boson $Z^0$
fluctuates into, with the nucleon target. The interaction is modeled
via the dipole-target cross section, whereas the boson fluctuation in
a color dipole is given by the corresponding light-cone wave
function. The neutral current (NC)  DIS structure functions
\cite{BGNPZ1,BGNPZ2,KUTAK} are related to the cross section for the scattering of transversely and longitudinally polarized $Z^0$ bosons. That is,
\begin{eqnarray}
F_{T,L}^{\mathrm{NC}}\,(x,Q^2) = \frac{Q^2}{4\,\pi^2\alpha_Z}\, \int d^2 \rr \,\int_0^1 dz \,
| \psi^{Z^0}_{T,L}\,(z,\,\rr,\,Q^2)|^2\,\sigma_{dip}\,(x,\,\rr)\,,
\label{FSDIP}
\end{eqnarray}
where  $\rr$ denotes the transverse size of the color dipole, $z$ is the
longitudinal momentum fraction carried by a quark and  $\psi^{Z}_{T,L}$ are  the light-cone wavefunctions for the virtual $Z^0$ bosons with transverse or longitudinal polarizations. The small-$x$ neutrino structure function $F_2^{\nu N}$ is computed from expressions above taking $F_2=F_T+F_L$. Explicit expressions for the light cone wavefunctions in the charged current case can be found at Refs. \cite{BGNPZ1,BGNPZ2}. The corresponding expressions for the special case of massless quarks and considering an isoscalar target are presented in Ref. \cite{KUTAK}. We quote Ref. \cite{gmm} and Refs. \cite{ZOLLER1,ZOLLER2} for the recent phenomenology using the color dipole picture in the case of CC neutrino interactions.

For the $Z^0$ boson, the coupling to quarks contains both vector and axial-vector parts and it is convenient to consider the basis of helicity spinors to decompose the $Z^0$ light-cone wave function into distinct vector and axial-vector parts \cite{MW}:
\begin{equation}
  \Psi_{\lambda} ^{\lambda_1 \lambda_2}(\vec{r},z,Q^2) =
  V_{\lambda} ^{\lambda_1 \lambda_2}(\vec{r},z,Q^2)
  - A_{\lambda} ^{\lambda_1 \lambda_2}(\vec{r},z,Q^2),
\end{equation}
where $\lambda_1,\lambda_2$ denote the quark helicities.  The probability densities of quark-antiquark states $|\Psi^{Z^0}_L|^2$ and $|\Psi^{Z^0}_T|^2$  are given by,
\begin{eqnarray}
|\Psi^{Z^0}_L|^2 & = & \sum_{\lambda_1,\lambda_2}\,\Psi_{0} ^{\lambda_1 \lambda_2}\left(\Psi_{0} ^{\lambda_1 \lambda_2}\right)^*=\sum_{\lambda_1,\lambda_2}\,\left(|V_{0} ^{\lambda_1 \lambda_2}|^2+ |A_{0}^{\lambda_1 \lambda_2}|^2 \right),\\
|\Psi^{Z^0}_T|^2 & = &\frac{1}{2} \left[\,|\Psi_{+1} ^{+1/2+1/2}|^2 +|\Psi_{+1} ^{-1/2+1/2}|^2 + |\Psi_{-1} ^{-1/2-1/2}|^2 + |\Psi_{-1} ^{-1/2+1/2}|^2 \, \right].
\end{eqnarray}

The vector and axial-vector parts of the $Z^0$ wave function are given by \cite{MW},
\begin{eqnarray}
  V^{\lambda_1 \lambda_2} _{\pm 1} & = &-\sqrt{\alpha_Z}g^f_V\sqrt{2N_c}\,
  \left\{\pm
  \mathrm{i}e^{\pm \mathrm{i}\theta_r}[
    z\delta_{\lambda_1,\pm}\delta_{\lambda_2,\mp} -
    (1-z)\delta_{\lambda_1,\mp}\delta_{\lambda_2,\pm}] \partial_r \, + \,   m_f \delta_{\lambda_1,\pm}\delta_{\lambda_2,\pm}
  \right\}\, \frac{K_0(\varepsilon r)}{2\pi},\nonumber\\
  V^{\lambda_1\lambda_2} _{0} & = & \sqrt{\alpha_Z} g^f_V
  \frac{\sqrt{N_c} }{Q}\,
  \delta_{\lambda_1,-\lambda_2} \, 2Q^2 z(1-z)\,
  \frac{K_0(\varepsilon r)}{2\pi}, \nonumber\\
  A^{\lambda_1 \lambda_2} _{\pm 1} & = & \sqrt{\alpha_Z} g^f_A \sqrt{2N_c}\,
  \left\{
  -\mathrm{i}e^{\pm\mathrm{i}\theta_r}[
    z\delta_{\lambda_1,\pm}\delta_{\lambda_2,\mp} +
    (1-z)\delta_{\lambda_1,\mp}\delta_{\lambda_2,\pm}] \partial_r \right.\nonumber \\
  & \pm & \left. m_f (1-2z) \delta_{\lambda_1,\pm}\delta_{\lambda_2,\pm}
  \right\}\, \frac{K_0(\varepsilon r)}{2\pi}, \nonumber\\
  A^{\lambda_1\lambda_2} _{0} & = & \sqrt{\alpha_Z}g^f_A \frac{\sqrt{N_c}}{Q}\,
  \left\{ \,
  \delta_{\lambda_1,-\lambda_2} \, 2\lambda_1\;
  \left[\,2Q^2 z(1-z) \, + 2m_f ^2\, \right] - \mathrm{i} \delta_{\lambda_1,\lambda_2} e^{-\mathrm{i}2\lambda_1\theta_r}
  2m_f\partial_r \;\right\}
  \frac{K_0(\varepsilon r)}{2\pi},\nonumber
\end{eqnarray}
where, $\alpha_Z = 4\pi\alpha_{em}/ \sin^2(2\theta_W)$ and $\varepsilon^2 = z(1-z)Q^2+m_f^2$. The axial-vector couplings are given by $g_V^f=\sqrt{\rho}\,(I_3^f-2Q^f\sin^2 \theta_W)$ and $g_A^f=\sqrt{\rho}\,(I_3^f)$. Here, $I_3^f$ and $Q^f$ are the weak isospin and electromagnetic charge of quark $f$, respectively. The relative coupling strength of the neutral to charged current interactions is labeled by $\rho$, where $\rho =1$ at tree level in SM. The weak mixing parameter, $\sin^2 \theta_W$, is related at tree level to $G_F$, $M_Z$ and $\alpha_{em}$ by $\sin^2 \theta_W = 4\pi\alpha_{em}/\sqrt{2}\,G_FM_Z^2$.

On the other hand, the dipole hadron cross section $\sigma_{dip}$  contains all
  information about the target and the strong interaction physics. Here, we consider two representative models for the dipole cross section.  In what follows one takes the phenomenological parameterizations: (a) Golec-Biernat-W\"{u}sthoff model (GBW) \cite{GBW} and  (b) impact parameter Color Glass Condensate model (b-CGC) model \cite{Watt}. Both models are able to describe experimental data on inclusive and diffractive deep inelastic $ep$ scattering at small-$x$. The b-CGC model is also successful in describing data on exclusive vector meson production at DESY-HERA and includes important aspects as impact parameter dependence, extended geometric scaling and saturation physics at very small-$x$.  The idea behind it is to introduce the impact parameter dependence into the CGC model \cite{Iancu:2003ge}, where dipole cross section is given by:
\begin{eqnarray}
\label{eq:bcgc}
 \sigma_{dip}(x,\,r)  & = & 2\,\int d^2b\,N\,(x,\,r,\,b), \\
 N\,(x,\,r,\,b) & = & \begin{cases}
  \mathcal{N}_0\left(\frac{rQ_{\mathrm{sat}}}{2}\right)^{2\left(\gamma_s+\frac{1}{\kappa\lambda Y}\ln\frac{2}{rQ_{\mathrm{sat}}}\right)} & :\quad rQ_{\mathrm{sat}}\le 2\\
  1-\mathrm{e}^{-A\ln^2(BrQ_{\mathrm{sat}})} & :\quad rQ_{\mathrm{sat}}>2
  \end{cases},
\end{eqnarray}
where $Y=\ln(1/x)$ and $Q_{\mathrm{sat}}(x,b)$ is the impact parameter saturation scale defined as \cite{Watt}:
\begin{equation}
\label{eq:bcgc1}
Q_{\mathrm{sat}}(x,b)=\left(\frac{x_0}{x}\right)^{\frac{\lambda}{2}}\;\left[\exp\left(-\frac{b^2}{2B_{\rm CGC}}\right)\right]^{\frac{1}{2\gamma_s}}.
\end{equation}
In the b-CGC model the evolution effects are included using an approximate solution to the Balitsky--Kovchegov equation. In particular, one has the slope $B_{\mathrm{CGC}}=5.5$ GeV$^{-2}$. The constants $A$ and $B$ are obtained from continuity conditions at  $rQ_{\mathrm{sat}}=2$.
We quote the original papers for details on the parameterizations and determination of their phenomenological parameters. Two important parameters are $\lambda$ and $\gamma_s$ (saturation anomalous dimension), which determine the energy dependence of the saturation scale $Q_{\mathrm{sat}}$.  Here, we use an effective light quark mass, $m_f=0.14$
GeV and the charm (bottom) mass is set to be $m_c=1.4$ GeV ($m_b=4.5$ GeV). In the dipole cross section we use $x=x_{Bj}[1+(4m_f^2/Q^2)]$.

For scattering on nuclei, we use the extension of the color approach for nuclear targets taking the Glauber-Gribov picture \cite{ARMESTO}, without any new parameter. In this
approach, the nuclear version is obtained replacing the
dipole-nucleon cross section  by the
nuclear one,
\begin{eqnarray}
\sigma_{dip}^{\mathrm{nucleus}} (x, \,r;\, A)  = 2\,\int d^2b \,
\left\{\, 1- \exp \left[-\frac{1}{2}\,T_A(b)\,\sigma_{dip}^{\mathrm{nucleon}} (x, \,r)  \right] \, \right\}\,,
\label{sigmanuc}
\end{eqnarray}
where $b$ is the impact parameter of the center of the dipole
relative to the center of the nucleus and the integrand gives the
total dipole-nucleus cross section for a  fixed impact parameter.
The nuclear profile function is labeled by $T_A(b)$ \cite{devries}.


\begin{figure}[t]
\centerline{\includegraphics[scale=0.55]{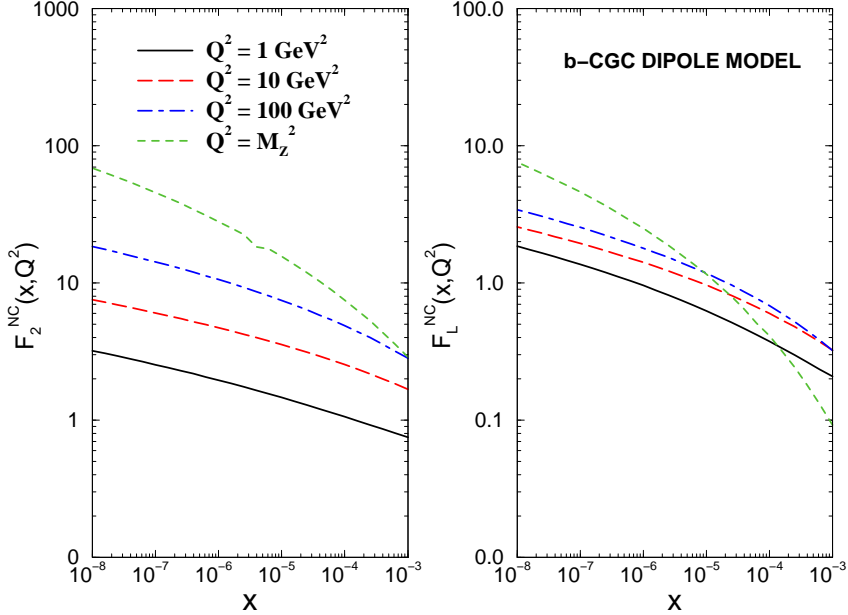}}
\caption{The structure functions $F_2^{NC}(x,Q^2)$  and $F_L^{NC}(x,Q^2)$ as a function of $x$ for distinct boson virtualities (see text) using the b-CGC model.}
\label{fig:1}
\end{figure}

In Fig. {\ref{fig:1}, the structure functions $F_2^{NC}(x,Q^2)$ (left panel) and $F_L^{NC}(x,Q^2)$ (right panel) are presented as a function of Bjorken variable $x$ for distinct boson virtualities using the b-CGC model. It is found that the $x$-dependence is approximately power-like with a effective power which growths on $Q^2$, $\lambda = \lambda (Q^2)$. Namely, it goes from $\lambda (Q^2=1 \,\mathrm{GeV}^2) \simeq 0.12$ up to $\lambda (Q^2= M_Z^2) \simeq 0.224$. Similar behavior is verified for the $F_L$ structure function, where a flattening is seen at very small-$x$. This is explained by the QCD evolution which is reproduced in the b-CGC model due to a running effective anomalous dimension. It is interesting to call attention to the unusual behavior of the structure function in the limit of large $Q^2$ and large $x$, which is more evident for $F_L$. In order to estimate the uncertainty from the theoretical side, we compute the structure function using the simple GBW saturation model, which is presented in Fig. \ref{fig:2}. The GBW model does not include the QCD evolution in the dipole cross section and in principle  it is not suitable for very large virtualities. However, the effective power is similar to the b-CGC model with $F_L$ being distinct at $Q^2=M_Z^2$ (it could be a consequence of missing QCD evolution in dipole cross section). It seems also that the flattening in $F_L$ is stronger in b-CGC than in GBW.


\begin{figure}[t]
\centerline{\includegraphics[scale=0.55]{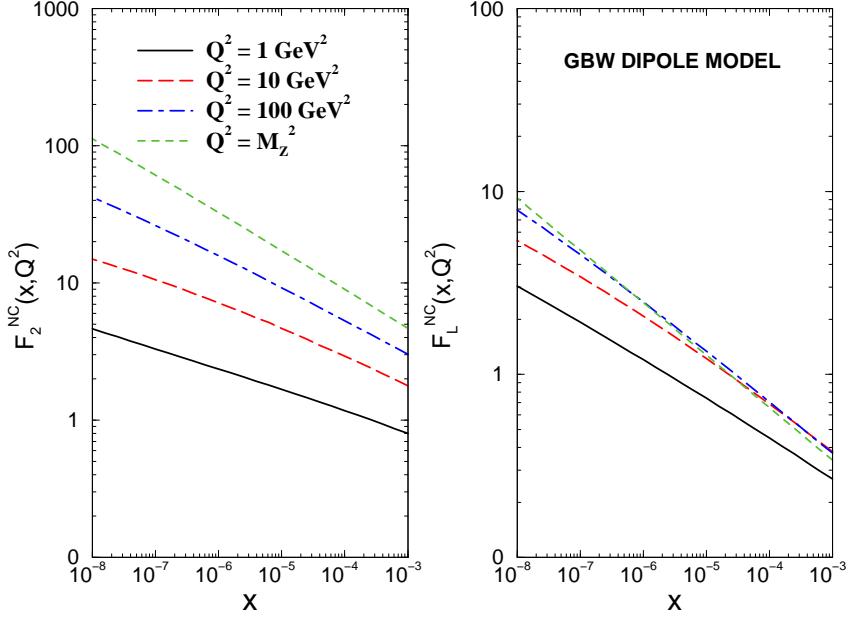}}
\caption{The structure functions $F_2^{NC}(x,Q^2)$  and $F_L^{NC}(x,Q^2)$ as a function of $x$ for distinct boson virtualities (see text) using the GBW model.}
\label{fig:2}
\end{figure}

In Fig. \ref{fig:3} we present the structure function $F_2$ as a function of $Q^2$ for fixed $x$. In left panel one has the virtuality dependence for b-CGC and GBW models for two representative values of $x$ ($x=10^{-2}$ and $x=10^{-6}$). It is verified the usual logarithmic dependence as the dipole models reproduce the scaling violation for structure functions. There is a small deviation on the overall normalization (a few percents), which is more sizable at large $Q^2$. In right panel, one has the flavor contribution to structure function. For sake of illustration we consider the GBW model and $x=10^{-3}$. For light quarks, $(d,\,s)$ contribution is dominant over $u$ quark, which is evident from the comparison among their electroweak couplings. The charm contribution corresponds to 13--14 \% of total result at this $x$ value. The bottom contribution is very small, suppressed by its larger mass.

An important feature of neutral current neutrino-nucleon interaction within the color dipole picture is the geometric scaling property. Such a scaling is a basic property of
the saturation physics. It means that  the total $\gamma^* p$
cross section at large energies is not a function of the two
independent variables $x$ and $Q$, but is rather a function of the
single variable $\tau_p = Q^2/Q_{\mathrm{sat}}^2(x)$ as shown
in Ref. \cite{SGK,PRLMAGVIC}. That is, $\sigma_{\gamma^*p}(x,Q^2)=\sigma_{\gamma^*p}(\tau_p)$. Similar scaling is predicted for lepton-nucleus scattering at high energies \cite{Armesto_scal}. In Ref. \cite{gmm} we have shown that scaling is present in charged current neutrino scattering. It is exact for light quarks, whereas is partially violated by charm contribution $c\bar{s}$ and $s\bar{c}$ due to different quark masses. We can qualitatively show the geometric scaling property in the NC neutrino-nucleon scattering by writing down the light-cone wavefunctions (for sake of simplicity, we take $g_V/g_A\approx 1$):


\begin{figure}[t]
\centerline{\includegraphics[scale=0.55]{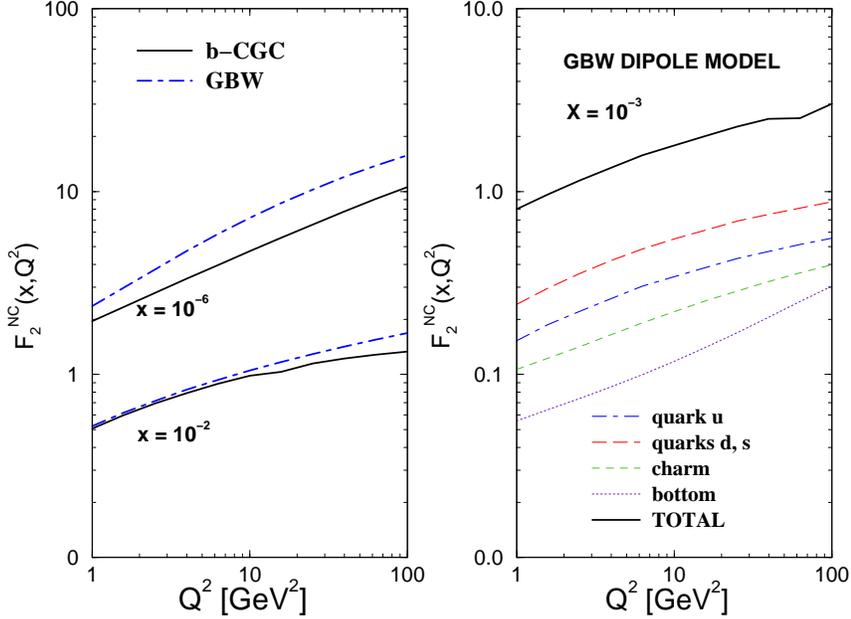}}
\caption{The structure function $F_2^{\nu N}(x,Q^2)$ as a function $Q^2$ for both models (left panel) and   its flavor contribution (right panel).}
\label{fig:3}
\end{figure}

\begin{eqnarray}
|\Psi_{T}^{Z^0}|^2  & \propto & \left[(1-z)^2+z^2 \right]m_f^2K_0^2(\varepsilon r)+ \left[(1-z)^2+z^2 \right]\varepsilon^2K_1^2(\varepsilon r),\\
|\Psi_L^{Z^0}|^2  & \propto & \left[4Q^4z^2(1-z)^2  +
|\left(2Q^2z(1-z)+2m_f^2\right)^2\right]\frac{K_0^2(\varepsilon r)}{Q^2}+ \left[ 4m_f^2\varepsilon^2\right]\frac{K_1^2(\varepsilon r)}{Q^2}. \nonumber \\
\end{eqnarray}

The equations above are different from the usual expressions for light-cone wavefunctions in electron-proton DIS even in the massless case due to the distinct weak couplings. Neglecting the quark mass, one can rewrite Eq. (\ref{FSDIP}) as,
\begin{eqnarray}
\sigma_{T,(L)}(Z^0+N\rightarrow X) & \propto & \int dz \int \bar{r}
d\bar{r}
f_{T,(L)}(z)K_{1,(0)}^2\left(\sqrt{z(1-z)}\,\bar{r}\right)\sigma_{dip}\left(\frac{Q_{\mathrm{sat}}^2}{Q^2}
\,\bar{r}\right),\nonumber \\
\end{eqnarray}
where we have $f_T(z) =[(1-z)^2+z^2](1-z)z$ and $f_L(z) =z^2(1-z)^2$. Here, we rescaled the transverse size $r$ to the dimensionless variable $\bar{r}=Qr$ and suppose a scaling behavior for the dipole cross section, $\sigma_{dip}=\sigma_{dip}(Q_{\mathrm{sat}} r)$. This last assumption is fullfiled by several saturation approaches, as for GBW model where  $\sigma_{dip}=\sigma_0[1-\exp(-r^2Q_{\mathrm{sat}}^2/4)]$.
 Therefore, we have a scaling appearing in the NC boson-nucleon cross section, $\sigma_{tot}(Z^0+N\rightarrow X)=\sigma_{tot}(\tau_p=Q_{\mathrm{sat}}^2/Q^2)$.
  Similar scaling should be present in the nuclear case, as shown in Ref. \cite{gmm}.

\begin{figure}[t]
\centerline{\includegraphics[scale=0.55]{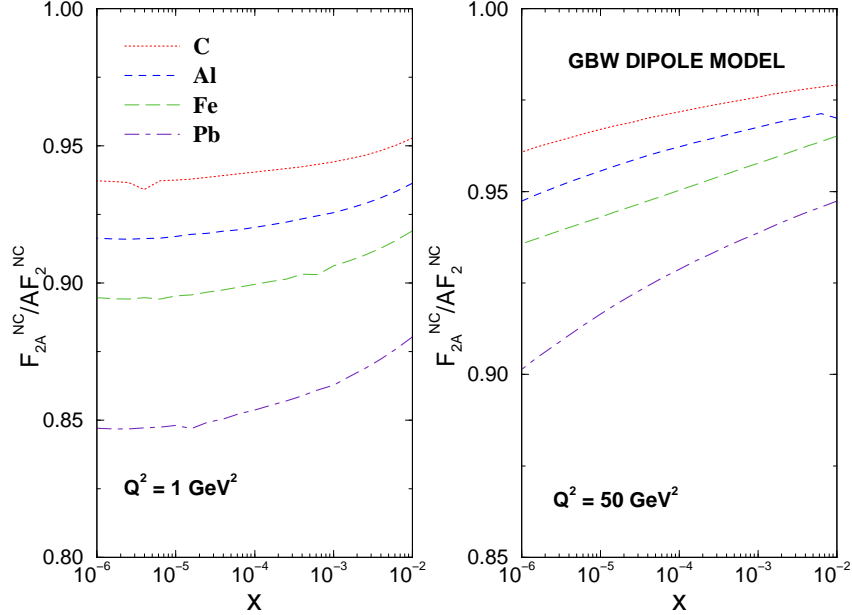}}
\caption{The nuclear ratio $R_A=F_{2A}^{NC}/AF_2^{NC}$ as a function of Bjorken variable for fixed virtualities ($Q^2=1$ and $50$ GeV$^2$) and several nuclear targets.}
\label{fig:4}
\end{figure}

In Fig. \ref{fig:4} we present the study for nuclear targets. The structure function $F_2^{NC}$ is presented as a function of $x$ for fixed values of virtualities ($Q^2=1$ GeV$^2$ and $Q^2=50$ GeV$^2$).
 We select the following nuclei: Carbon, Aluminum, Iron and Lead. We stress that for NuSonG experiment we will have an isoscalar target similar to $A=Fe$ as for NuTeV. The nuclear shadowing is more intense for heavy nuclei, smaller $x$ and virtualities. In particular, for Iron targets the nuclear shadowing is of order 10\% at $Q^2=1$ GeV$^2$, whereas it reaches to 15--20 \% for Lead. Nuclear shadowing is small at large virtualities even at low-$x$. This fact is also verified in charged current interactions as investigated in Ref. \cite{gmm}, where the nuclear ratio has been computed for $F_2^{CC}$ and $xF_3$.
Some questions are in order here. The information on isospin violation
 is quite relevant for electroweak
measurements and isospin symmetry is assumed in the NuTeV analysis
 \cite{Tzanov}. Violation of isospin symmetry can be one important
 source of NuTeV anomaly. Therefore, better constraint on  isospin violation will be crucial to the interpretation of new neutrino experiments. The NuSOnG experiment \cite{nusong} purposes to  measure separately  $\Delta F_{2}\equiv\frac{5}{18}\, F_{2}^{CC}(x,Q^{2})-F_{2}^{NC}(x,Q^{2})$
in CC $W^{\pm}$ exchange and NC
$\gamma/Z$ exchange processes, which can also constrain charge symmetry violation. As NuSOnG (glass target, ${\rm SiO}_{2}$, is very nearly isoscalar) could measure $F_{2}^{CC}$ on a variety of targets, this will reduce the systematics associated with the heavy nuclear
target corrections providing new information on isospin violation. In particular,
for ${\rm SiO}_{2}$ we have $Z_{\rm O}=8,$ $Z_{\rm Si}=14$, $m_{\rm O}=15.994$, $m_{\rm Si}=28.0855$.
Using $A=Z+N$ we have $(N-Z)/A=(A-2Z)/A$, which yields $(N-Z)/A\sim-0.000375$ for $O$ and $(N-Z)/A\sim0.00304$
for Si. Therefore, since NuSOnG will provide high statistics
$F_{2}^{CC}$ measurements for a variety of $A$ targets, this will
yield an alternate handle on the isospin violation and will also improve our understanding
of the associated nuclear corrections.

As a final study, we compute the associated charm production in NC neutrino-nucleus interactions, $\sigma_{\mathrm{NC}} (c\bar{c})/\sigma_{\mathrm{NC}}^{DIS}$. The total NC neutrino-nucleon cross sections as a function of the neutrino energy and atomic number are given by the integration over available phase space and read as,
\begin{eqnarray}
\sigma^{\mathrm{NC}}_{(\nu,\,\bar{\nu})}(E_{\nu};\,A)=\int _{Q_{\mathrm{min}}^2}^{s} \! dQ^2\int_{Q^2/s}^1 \! dx \,\frac{1}{xs}
\frac{\partial^2 \,\sigma_{(\nu,\,\bar{\nu})}^{\mathrm{NC}}}{\partial
  x\,\partial y}, \label{signutotal}
\end{eqnarray}
where
\begin{eqnarray}
\frac{\partial^2 \,\sigma_{(\nu,\,\bar{\nu})}^{\mathrm{NC}}}{\partial x\,\partial y} = \frac{G_F^2\,m_N \,E_{\nu}}{\pi}\left(\frac{m_{Z}^2}{Q^2+m_{Z}^2}\right)^2 \left[\frac{1+(1-y)^2}{2} F_2^{\mathrm{NC}}(x,Q^2;\,A) - \frac{y^2}{2}F_L^{\mathrm{NC}}(x,Q^2;\,A)\right], \nonumber
\label{difxsecnu}
\end{eqnarray}
where $G_F$ is the weak Fermi coupling constant, $m_N$ is the nucleon mass, $E_{\nu}$ is the incident neutrino energy, $Q^2$ is the square of the four-momentum transfer to the nucleon. The variable $y=E_{had}/E_{\nu}$ is the fractional energy transferred to the hadronic vertex with $E_{had}$ being the measured hadronic energy, and $x=Q^2/2m_NE_{\nu}y$ is the Bjorken scaling variable (fractional momentum carried by the struck quark). The minimum $Q_{\mathrm{min}}^2 \propto O(1)$ GeV$^2$ is introduced in order to stay in the DIS region. Here, one considers UHE neutrinos, where the valence quark contribution stays constant  and physics is driven  by sea quark contributions. Hence, the $xF_3$ contribution should be negligible and it will be disregarded. Similar cross section can be computed for the charm contribution by replacing the respective charm structure function. We quote Refs. \cite{MPRD1,MPRD2} for previous analysis on CC and NC neutrino cross sections using the color dipole picture.

Recently, the CHORUS Collaboration \cite{CHORUS} has reported an extraction of the associated charm production  in NC neutrino DIS at $E_{\nu}=27$ GeV. The experimental determination found $\sigma(c\bar{c})/\sigma_{tot}= 3.62^{+2.95}_{-2.42}\,(\mathrm{stat})\pm 0.54\,(\mathrm{syst})\times 10^{-3}$ at the neutrino energy $E_{\nu}=27$ GeV. We quote also previous determinations, as the E531 experiment \cite{E531} $\sigma(c\bar{c})/\sigma_{tot}= 1.3^{+3.1}_{-1.1}\times 10^{-3}$  for energy $E_{\nu}=22$ GeV and the NuTeV experiment \cite{NuTeV} giving $\sigma(c\bar{c})/\sigma_{tot}= 6.4^{+5.5}_{-4.6}\times 10^{-3}$ at $E_{\nu}=154$ GeV. We have found theoretically $\sigma(c\bar{c})/\sigma_{tot}= 2.7\times 10^{-2}$ for $E_{\nu}=27$ GeV and $\sigma(c\bar{c})/\sigma_{tot}= 1.35\times 10^{-1}$ for $E_{\nu}= 154$ GeV. Our results are systematically above the experimental measurements, which can be explained by the extrapolation to very low energies of the color dipole results for charm contribution. It is well know that the dipole approaches overestimate the structure functions for $x\rightarrow 1$. At high energies our predictions are consistent with other approaches using small-$x$ formalisms.

\section{Comments and Conclusions}

As a summary, an analysis of small-$x$ neutral current (NC)
neutrino-nucleus DIS is performed within the color dipole
formalism. The structure functions $F_2^{\nu N}$,  $F_L^{\nu N}$ and
the nuclear ratio $R_A = F_2^{\nu A}/AF_2^{\nu N}$ are calculated. In
order to investigate the theoretical uncertainties we employed two
phenomenological parameterizations for the dipole cross section which
successfully describe small-$x$ inclusive, diffractive $ep$ DIS data
and exclusive production of vector mesons.  Nuclear shadowing is
taking into account through Glauber-Gribov formalism.  We have found
deviations among the models at very small-$x$ and mostly at large
virtualities. It is found that small-$x$ data show geometric scaling
property for the NC boson-hadron cross section as a function of the
scaling variable $\tau$. Although the results presented here are
compelling, further investigations are requested. In particular, it
would be useful compare the present calculations to measurements of
neutrino-nucleus structure function in smaller values of $x$ than the
currently measured in the accelerator experiments. This is the reason
we have focused on the NuSonG purpose, taht  will provide high statistics
$F_{2}^{CC}$ and $F_2^{NC}$ measurements for a variety of nuclear
targets. We also compute the charm content to the total NC neutrino
cross section and we show that it is consistent with current
experimental measurements. So, we have a description of small-$x$
structure functions in neutrino-proton interactions using a color
dipole formalism. Despite the limitations of the current approach for
the low energy neutrinos, our estimates are comparable to NC
neutrino-nuclei data for charm production. Our calculation is also
consistent with other approaches in literature in the high energy limit.

\section*{Acknowledgments}

This work was supported by CNPq (Brazil).

\ed
\begin{thebibliography}{99}

\bibitem{gmm}
M.B. Gay Ducati, M.M. Machado and M.V.T. Machado, Phys. Lett. B {\bf 644 }, 340 (2007).

\bibitem{DIPOLEPIC}
A. H. Mueller,  Nucl. Phys. {\bf B335} (1990) 115;
N.N. Nikolaev and B.G. Zakharov,  Z. Phys. {\bf C49}
(1991) 607.


\bibitem{CCFR}
W.G. Seligman {\it et al.} [CCFR Coll.], Phys. Rev. Lett. {\bf 79} (1997) 1213.

\bibitem{Fleming}
B.T. Fleming {\it et al.} [CCFR Coll.],  Phys. Rev. Lett. {\bf 86} (2001) 5430.

\bibitem{CCFR3}
U.K. Yang {\it et al.},  Phys. Rev. Lett. {\bf 86} (2001) 2742.

\bibitem{Tzanov}
M. Tzanov {\it et al.} [NuTeV Coll.],  Phys. Rev. {\bf D74} (2006) 012008.

\bibitem{nusong} T. Adams {\it et al.} [NuSOnG Coll.], arXiv:0803.0354
  [hep-ph], J.M. Conrad, AIP Conf. Proc. {\bf 981}, 243 (2008).

\bibitem{PW}  E.~A.~Paschos and L.~Wolfenstein,
  Phys.\ Rev.\  D {\bf 7}, 91 (1973).

\bibitem{ARMESTO}
N. Armesto, Eur. Phys. J. {\bf C26} (2002) 35.

\bibitem{BGNPZ1}
V. Barone, M. Genovese, N.N. Nikolaev, E. Predazzi and B.G. Zakharov
 Phys.Lett. {\bf B292} (1992) 181.

\bibitem{BGNPZ2}
V. Barone, M. Genovese, N.N. Nikolaev, E. Predazzi and B.G. Zakharov
 Phys.Lett. {\bf B328} (1994) 143.
\bibitem{KUTAK} K. Kutak and J. Kwieci\'nski,  Eur. Phys. J. C {\bf 29}, 521 (2003).

\bibitem{ZOLLER1} R. Fiore and V.R. Zoller, JETP Lett. {\bf 82} (2005) 385.

\bibitem{ZOLLER2} R. Fiore and V.R. Zoller, Phys. Lett. {\bf B632} (2006) 87.

\bibitem{MW} L. Motyka and  G. Watt, Phys. Rev. {\bf D78} (2008) 014023.

\bibitem{GBW} K. Golec-Biernat and  M. W\"usthoff,   Phys. Rev. {\bf D59} (1999) 014017,  {\it ibid.} {\bf D60} (1999) 114023.

\bibitem{Watt} G. Watt and H. Kowalski, Phys. Rev. {\bf D78} (2008) 014016.

\bibitem{Iancu:2003ge}
  E.~Iancu, K.~Itakura and S.~Munier,
  Phys.\ Lett.\ B {\bf 590} (2004) 199.

\bibitem{devries}
C. W. De Jager, H. De Vries, C. De Vries, Atom. Data Nucl. Data Tabl. {\bf 14}, 479 (1974).


\bibitem{SGK} A.M.  Sta\'sto, K. Golec-Biernat and J. Kwieci\'nski,   Phys. Rev. Lett. {\bf 86} (2001) 596.

\bibitem{PRLMAGVIC}
  V.~P.~Goncalves and M.~V.~T.~Machado,
  Phys.\ Rev.\ Lett.\  {\bf 91}, 202002 (2003).

\bibitem{MPRD1} M.V.T. Machado, Phys. Rev. {\bf D71} (2005) 114009.

\bibitem{Armesto_scal}
  N.~Armesto, C.~A.~Salgado and U.~A.~Wiedemann,
  Phys.\ Rev.\ Lett.\  {\bf 94}, 022002 (2005).
\bibitem{MPRD2} M.V.T. Machado, Phys. Rev. {\bf D70} (2004) 053008.

\bibitem{CHORUS} A. Kayis-Topaksu {\it et al.} [CHORUS Coll.], Eur. Phys. J. {\bf C52} (2007) 543.

\bibitem{E531} N. Ushida {\it et al.} [E531 Coll.], Phys.\ Lett.\ B {\bf 206} (1988) 375.

\bibitem{NuTeV} M. Goncharov {\it et al.} [NuTeV Coll.], Phys. Rev. {\bf D64} (2001) 112006.


\end{thebibliography}
